\documentclass[twocolumn,twocolappendix]{aastex631}

\usepackage{CJK}
\usepackage[dvipsnames]{xcolor}
\usepackage{hyperref}
\usepackage{subfigure}

\begin{document}
\begin{CJK*}{UTF8}{gbsn}

\title{The Pre-Outburst Properties of the FU Ori Object HBC 722}

\author[0000-0002-7154-6065]{Gregory J. Herczeg (沈雷歌)}
\affiliation{Kavli Institute for Astronomy and Astrophysics, Peking University, Beijing 100871, China, gherczeg1@gmail.com}
\affiliation{Department of Astronomy, Peking University, Beijing 100871, China}

\author[0000-0001-8174-1932]{Bo Reipurth}
\affiliation{Institute for Astronomy, University of Hawaii in Manoa, 640 North Aohoku Pl., Hilo, HI 96720, USA}
\affiliation{Planetary Science Institute, 1700 E Fort Lowell Rd Ste 106, Tucson, AZ 85719, USA}

\begin{abstract}
FU Ori outbursts are thought to play an important role in stellar assembly and the evolution of protoplanetary disks.  However, the progenitor young stellar objects are largely uncharacterized.  We obtained a low-resolution optical spectrum of HBC 722 before its FU Ori outburst as part of a survey of young stellar objects in the North America Nebula.  The spectrum yields a spectral type of M3.3$\pm$0.4, which when combined with archival photometry allows us to measure the stellar and accretion properties of a young star prior to its FU Ori outburst.  The pre-outburst accretion rate of $7\times10^{-9}$ M$_\odot$ yr$^{-1}$ is high for a protoplanetary disk around an M3-M3.5 star, though about 15,000 times weaker than the accretion rate during the outburst.  
The pre-outburst variability, inferred from archival B-band photometry, is about a factor 5 with a standard deviation of 0.16 dex and is consistent with variable accretion onto young low-mass stars.
The stellar radius is larger than the radius of accreting young stars of similar spectral type by a factor of two.  
The extinction to HBC 722 is $\sim 1.45\pm0.3$~mag, lower than the 2.5--3.7~mag extinction values measured during the outburst.  The u-band photometry plays an especially important role in constraining the veiling at longer wavelengths and therefore also the extinction and photospheric luminosity.  
\end{abstract}


\section{Introduction} \label{sec:intro}

FU~Orionis objects (FUors) were recognized as major events occurring in pre-main sequence objects by \citet{herbig66} and \citet{herbig77}.
 During large eruptions of up to 6~mag in amplitude, FUors display distinct spectroscopic features, including strong CO absorption, weak metal absorption, strong water bands, low gravity, strong P~Cyg profiles in specific lines, and at most a few emission lines \citep[e.g.][]{connelley18,rodriguez22}.  These combined features are interpreted as a viscously heated accretion disk with accretion rates of $\sim 10^{-4}$ to $10^{-5}$\,M$_\odot$ yr$^{-1}$ \citep[e.g.][]{popham96}, 3-4 orders of magnitude larger than accretion onto typical T Tauri stars \citep[see reviews by][]{hartmann96,hartmann16,manara23}. These massive accretion events likely play a crucial role in stellar mass assembly and in the chemical evolution of disks and envelopes \citep[see review by][]{fischer23}, though the triggering mechanism remains uncertain (see for example discussions in \citealt{audard14} and \citealt{vorobyov21}).

FU Ori outbursts are rare. Many of the classical and even recently discovered outbursts began prior to the modern era of systematic spectroscopic surveys \citep[see for instance the FU Ori-like objects V646~Pup, \citealt{reipurth02}, and RNO 54,][]{hillenbrand23}.  In the era of time domain astronomy, the discovery of FU Ori objects has proliferated \citep[see summary by][]{contreras25oy}, but most are distant and previously anonymous objects that lack pre-outburst spectra.   As a consequence, the stellar and accretion properties of the progenitors of FUors are not well constrained.

\begin{figure*}[!t]
\centering
\includegraphics[trim = 50mm 31mm 25mm 133mm, width=0.85\textwidth]{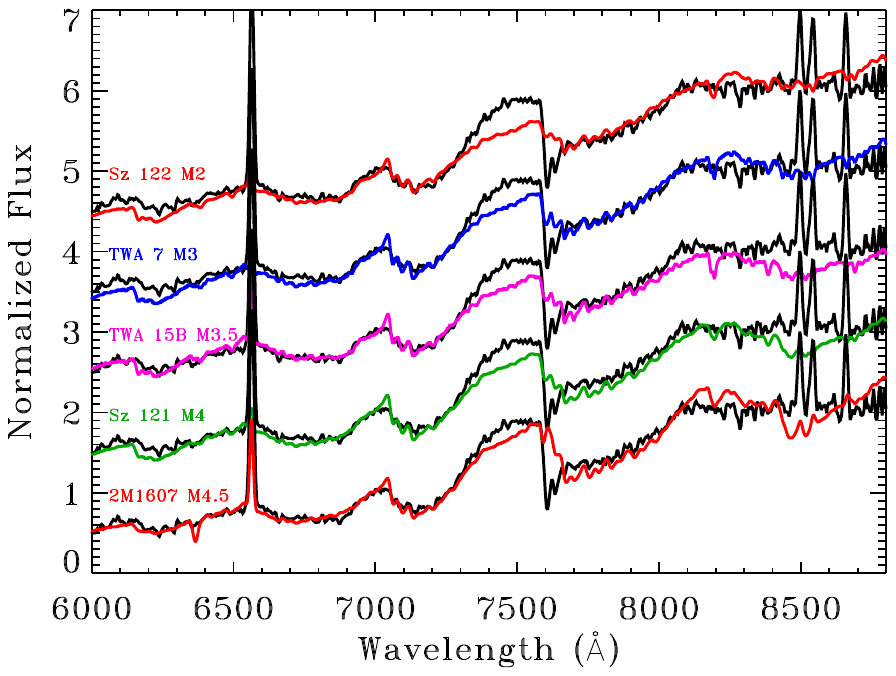}
\caption{The WFGS2 red spectrum of HBC 722 compared with spectra of non-accreting X-Shooter templates from \citet{claes24}.  The templates all have veiling and extinction tailored to best match the observed spectrum of HBC 722.  The spectrum also includes H$\alpha$ and \ion{Ca}{2} infrared triplet lines, indicating accretion.
}
\label{fig:spectfit}
\end{figure*}

In this context, HBC 722 (LkH$\alpha$~188~G4, V2493~Cyg, PTF 10qpf) is a still-unique case of a recent FU Ori burst with a pre-outburst spectrum with high enough quality to measure stellar and accretion properties.  HBC~722 erupted and brightened by $\sim$4 magnitudes between July 2009 and May 2010 \citep{semkov10a,semkov10b,semkov10c}. Extensive optical and near-infrared spectroscopy revealed a classical FU~Ori-type spectrum (see detailed studies by \citealt{miller11} and \citealt{carvalho24}, along with a review of this outburst by \citealt{semkov21}).

Little was known about HBC 722 before its outburst.  The star was originally discovered by \citet{herbig58} in a survey for H$\alpha$ emission stars, including the little group LkH$\alpha$ 185-189, in the Cygnus star-forming region.  \citet{ck79} noted another five faint H$\alpha$ emission stars associated with the group and labeled them as LkH$\alpha$~188 G1 to G5. HBC 722 belongs to the L935 cloud, which forms the 'Gulf of Mexico' and separates the North America and Pelican Nebulae.  Later, \citet{armond11} identified in L935 an additional 30 new H$\alpha$ emission stars and a large group of 35 new Herbig-Haro objects, testifying to the youth of the region.
Pre-outburst spectra exist because of its location and H$\alpha$ emission \citep{ck79,semkov12,fang20}.

In this paper, we reassess the pre-outburst stellar and accretion properties of HBC 722 by analyzing our previously unpublished low-resolution optical spectrum and re-interpreting pre-outburst photometry.  Our results, including the spectral type, are consistent with but improve upon measurements in pre-outburst spectra analyzed by \citet{ck79}, \citet{fang20}, and a dedicated analysis by \citet{carvalho24}, primarily because our spectrum is redder and therefore provides a more accurate spectral type than previously published spectra. 
In \S 2 we present the observations used in this paper.  In \S 3 we measure the contributions of the photosphere and accretion to photometry, adjusted for extinction, in \S 4 we convert those measured properties to stellar and accretion properties, and in \S 5 we discuss the implications of our analysis for understanding the increase in accretion and likely increase in extinction during the FU Ori outburst.

\section{Observations} \label{sec:obs}

We obtained a low-resolution red spectrum of HBC~722 on 2002 July 14/15 using the Wide Field Grism Spectrograph-2 \citep[WFGS2, ][]{uehara04}, installed at the University of Hawaii 2.2m telescope at Mauna Kea. A red grism was used with a 0.9 arcsec wide slit, providing a spectral coverage from 6000--9000 \AA\ with a resolution of $\sim 1000$. A total of three 12 min exposures were obtained in seeing of 1.1~arcsec in photometric conditions. The standard Feige~110 was observed at the same airmass, although the spectrum of HBC 722 is not accurately flux calibrated.

We also use archival optical and infrared photometry to re-evaluate the star and accretion properties of HBC 722.  The SDSS ugriz photometry \citep[DR12, PSF magnitudes]{alam15} was obtained  within 300\,s for all bands, minimizing variability.  In addition, \citet{semkov12} provides BVRI photometry before the outburst on 80 nights, with some bands missing on some nights (hereafter referred to as Semkov photometry).  On 2007 Aug 16 (MJD 54329), the B-band brightness was 16.6, about 3 mag brighter than normal, despite VRI brightness consistent with the bright end of the light curves.  This B-band outlier is considered unreliable and is excluded from all analyses in this paper.
We supplement the optical photometry with JHK$_{\rm S}$ photometry from 2MASS \citep{cutri03}, JHK photometry from UKIDSS \citep{lucas08}, and Spitzer/IRAC mid-IR emission from \citet{guieu09}.

In \S 3, we model observed photometry with a photosphere, accretion continuum, and dust disk emission. For these fits, uncertainties in the photometry are crudely assessed as 0.03 mag in most bands, 0.06 mag in g and B, and 0.37 mag in u.
The uncertainties in these bands are intended to include changes due to short-timescale variability ($\sim 0.01$ in TESS, \citealt{robinson22}, \citealt{zsidi22}, and Ji et al.~submitted) and errors in convolving flux-calibrated spectra to photometric systems, both of which dominate over the measurement errors (for sdss, $\sim 0.03$ in griz).  
The SDSS u-band PSF photometry is $21.09\pm0.37$, which is very faint, but data release notes indicate that the u-band image has a profile consistent with a star.  The uncertainty is treated as symmetric, though it may be lopsided and larger towards fainter magnitudes.
In model fits to photometry, the uncertainty in JHK photometry is multiplied by $\sqrt{2}$ to avoid double-counting the significance of non-independent points.

Filter curves are obtained from the Spanish Virtual Observatory \citep{rodrigo12}.  
The Semkov photometry is modeled with Johnson BV and Cousins RI transmission curves and zero points.  Some offsets between our synthetic photometry and the Semkov photometry are apparent in R and I and are discussed further in \S 3.3 and Figure~\ref{fig:cmds}.

\section{Empirical fits to spectra and photometry}

\begin{figure*}[!t]
\centering
\includegraphics[trim = 30mm 133mm 30mm 30mm, width=0.85\textwidth]{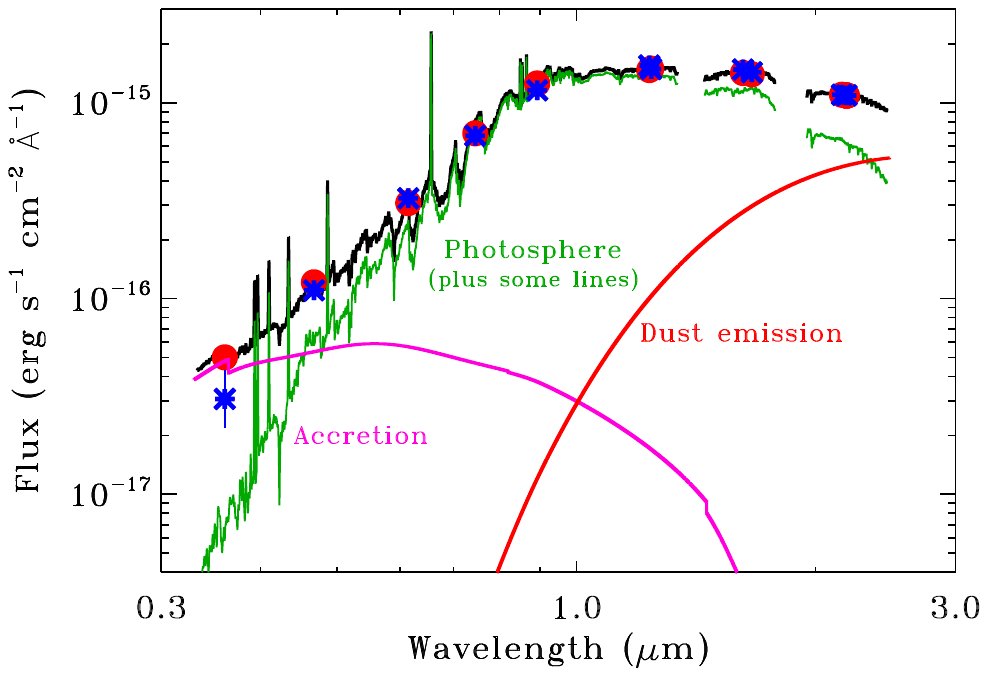}
\caption{The synthetic photometry (red circles, calculated from the synthetic spectrum) that best reproduces the SDSS griz, 2MASS JHK$_{\rm S}$, and UKIRT JHK photometry of HBC 722 (blue asterisks).  The synthetic spectrum (black line with red circles for photometry) consists of an M3.25 photosphere plus selected emission lines (green line), an accretion continuum (purple line), and a dust continuum emission (red line).}
\label{fig:sedplot}
\end{figure*}

\subsection{Spectral Classification of HBC 722}

The red spectrum of HBC 722 is typical of an M-type classical T Tauri star, with deep and broad molecular absorption bands (primarily TiO) from the photosphere and strong H$\alpha$ and \ion{Ca}{2} infrared triplet emission from the accretion flow.  No other lines are detected.

Figure~\ref{fig:spectfit} compares the spectrum of HBC 722 to spectra of X-Shooter templates from FRAPPE (\citealt{claes24}, adapted from \citealt{manara13} and \citealt{manara17}), with veiling and extinction assessed to best match the observed spectrum at each spectral type.  We calculate best fits by comparing the observed spectrum to each FRAPPE spectrum, with free parameters of the photospheric temperature, extinction (\citealt{cardelli89} with $A_V=3.1$), veiling \citep[assumed to be flat in this spectral analysis,][]{herczeg14}, and a final scaling.
The best fit from the FRAPPE templates leads to an M$3.3\pm0.4$ spectral type.  The corresponding temperature of $3350\pm75$ K is obtained from \citet{herczeg14} and is consistent with temperatures of FRAPPE templates and the analysis by \citet{fang25}.
The veiling in the spectral range between 6600--6700 \AA\ is $<0.2$.  
Because the flux is not calibrated, the extinction (best fit of $A_V<1$ mag) serves as an artificial way to adjust the broad continuum shape and is not considered reliable.  These values are similar to the results obtained by applying the spectral grid from \citet{herczeg14}.

The equivalent widths for emission lines are 60 \AA\ for H$\alpha$ and 7.8, 7.4, and 6.3 \AA\ for \ion{Ca}{2} 8498, 8448, 8662 \AA.  The \ion{He}{1} 6678 \AA\ line is not detected, with a 2$\sigma$ upper limit of 1.5 \AA.

\subsection{Photosphere and accretion parameters from SDSS photometry}

We reproduce the SDSS photometry by calculating synthetic photometry from 
a spectrum that includes a template photosphere, an accretion continuum, and dust continuum emission (Figure~\ref{fig:sedplot}).  The template photosphere is fixed to approximate an M3.25 star, a combination of the X-shooter spectra of the M3 star TWA 7 and the M3.5 star TWA 15B \citep[from FRAPPE,][]{claes24}, both scaled to have equal flux at 7000\,\AA.  The accretion continuum is approximated as a pure hydrogen slab, as calculated by \citet{valenti93}.  
The dust continuum is produced by a disk with a temperature maximum of  1400 K and temperature that depends inversely on radius, $r^{-1}$, with a slope chosen to roughly reproduce the 2MASS Ks and IRAC photometry.  
H$\alpha$ and \ion{Ca}{2} IR triplet lines are added, with equivalent widths as measured from the low-resolution spectrum.  Higher Balmer lines and the \ion{Ca}{2} H \& K lines are also added with similar equivalent widths, fixed to the photospheric emission and not scaled with accretion.  
The free parameters are the extinction \citep{cardelli89} and scaling factors for the accretion continuum and photosphere.

The synthetic photometry is then fit to the SDSS griz and the 2MASS and UKIDSS JH photometry.  The dust continuum is fixed to exactly recover the K-band magnitude and serves to contribute to the J and H bands.  When the synthetic u-band photometry is fainter than observed, it is automatically scaled up to the observed photometry, consistent with a larger Balmer Jump.  Thus, the u-band only contributes to the goodness of fit if the synthetic emission is brighter than the observed emission.

\begin{figure*}[!t]
\centering
\includegraphics[trim = 30mm 135mm 30mm 30mm, width=0.45\textwidth]{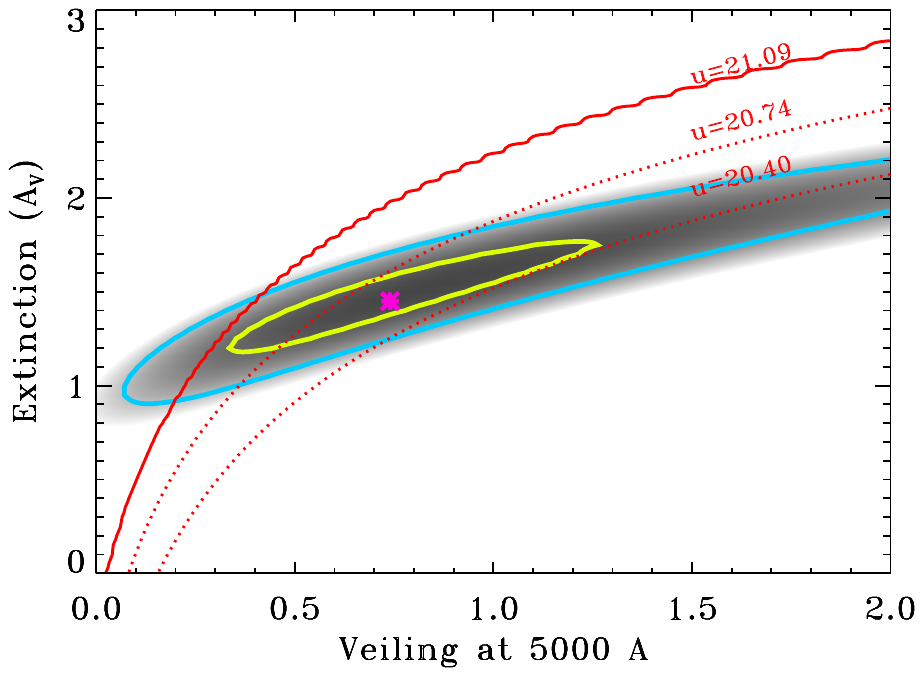}
\includegraphics[trim = 30mm 135mm 30mm 30mm, width=0.45\textwidth]{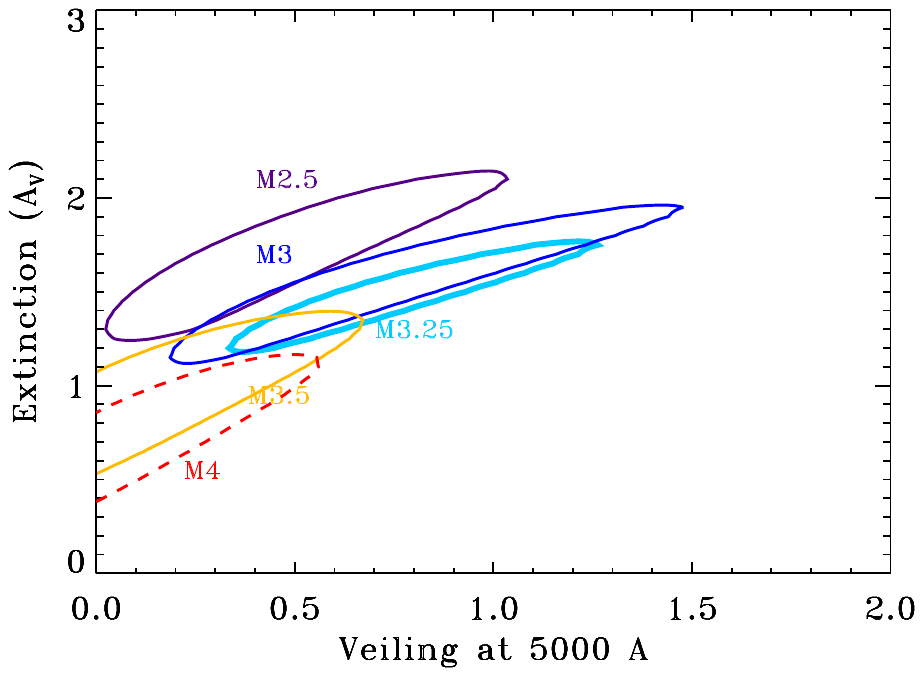}
\caption{Left:  Contour plot of the acceptable parameters of veiling and extinction for synthetic photometry fit to SDSS photometry.  The best fit (purple asterisk) is shown in Figure~\ref{fig:sedplot}.  
 The yellow and blue contours correspond to the acceptable fit range, with $\chi_{\rm red}^2=1.0$ and $1.5$.  
The broader region shows the $\chi^2$ contours for fits to griz and excluding the u-band, with acceptable parameters that extend to high veiling.  Those values are ruled out because the synthetic spectrum with higher veiling and extinction produces a u-band brightness that is brighter than detected (red contours for $u=21.0$, the $1-\sigma$ limit, and 20.5, which is 0.5 mag brighter than observed).  Right: contours of $\chi_{\rm red}^2=1.0$ for spectral types of M2.5 (purple), M3 (Sz 67, dark blue), M3.25 (TWA 7+TWA 15B, thick light blue), M3.5 (TWA 15B, orange), and Sz 121 (red dotted line indicating $\chi_{red}^2=1.7$, since no solutions were within $\chi_{red}^2=1.3$).  In general the extinction and veiling both decrease to later spectral type, but no solutions reach $A_V>2$.}
\label{fig:contour}
\end{figure*}

Figure~\ref{fig:contour} shows the acceptable contours for $\chi_{red}^2=1$ and $1.5$ for veiling and extinction for the M3.3 photosphere.  The fits have 4 degrees of freedom, though the assigned photometric errors are somewhat arbitrary.
Without the u-band, the acceptable parameter space extends to high veiling and larger extinctions.  However, those parameters lead to u-band photometry that is significantly larger than the observed value, even with a low value for the Balmer Jump.  The low veiling solution is also consistent with the red optical spectrum. 
The spectral fit rules out any veiling $>0.2$ at 6600--6700 \AA, which corresponds to $>0.85$ at 5000 \AA, with the caveat that the photometry at the exact time of the low-resolution spectrum is not known. Our adopted best fit is near this limit.

Fits for photospheric templates for spectral types M2.5--M4 lead to similar contours (right panel of Fig.~\ref{fig:contour}).  None of the fits produce an extinction above $>1.8$.  The best-fit parameters tend to zero veiling at 5000 \AA\ for later spectral types, which is inconsistent with the accretion variability seen in the Semkov photometry (see \S 3.3).  Moving the infrared photometry together upward and downward by 0.1 mag leads to only minor changes in the acceptable contours.

These fits assume the shape of the accretion continuum spectrum (see discussions in, for example, \citealt{herczeg23} and \citealt{pittman25}), the contribution from emission lines, extinction curve, and spot coverage.  Addressing these uncertainties is beyond the scope of this paper and is challenging to assess without multi-epoch, flux-calibrated spectroscopy.

\subsection{Accretion variability from Semkov photometry}

\begin{figure*}[!t]
\centering
\includegraphics[trim = 30mm 123mm 30mm 30mm, width=0.8\textwidth]{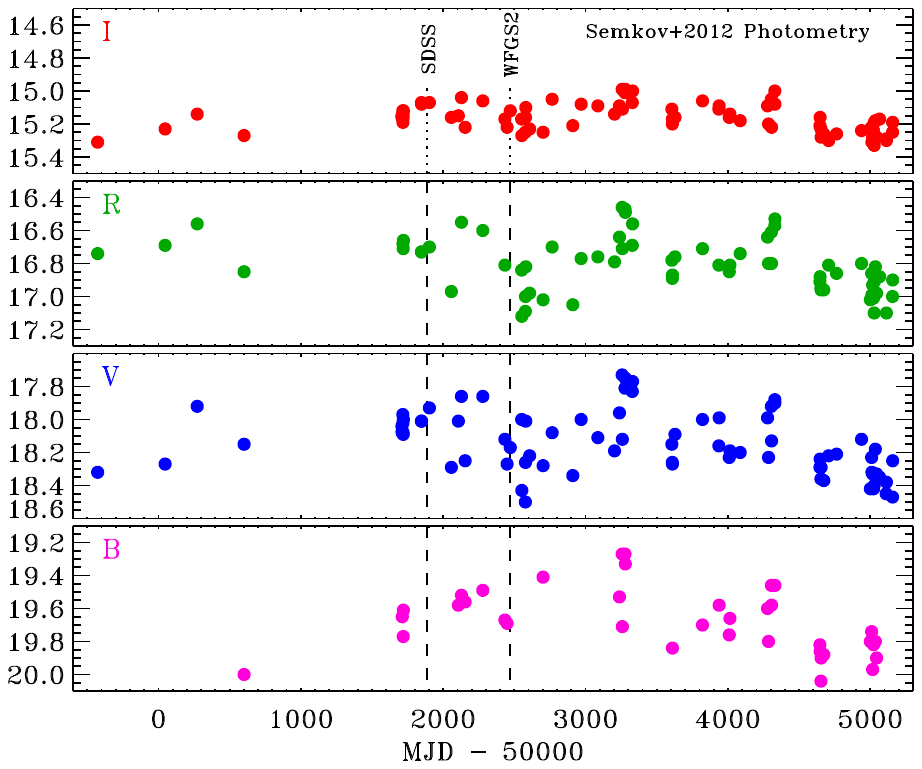}
\caption{The pre-outburst BVRI light curve of HBC 722 from \citet{semkov12}, with vertical dashed lines showing the epoch of the SDSS photometry and the WFGS2 spectrum.  The outlier $B$-band point is excluded from this Figure and Figure~\ref{fig:cmds}.}
\label{fig:lightcurve}
\end{figure*}

\begin{figure*}[!t]
\centering
\includegraphics[trim = 20mm 133mm 30mm 30mm, width=0.32\textwidth]{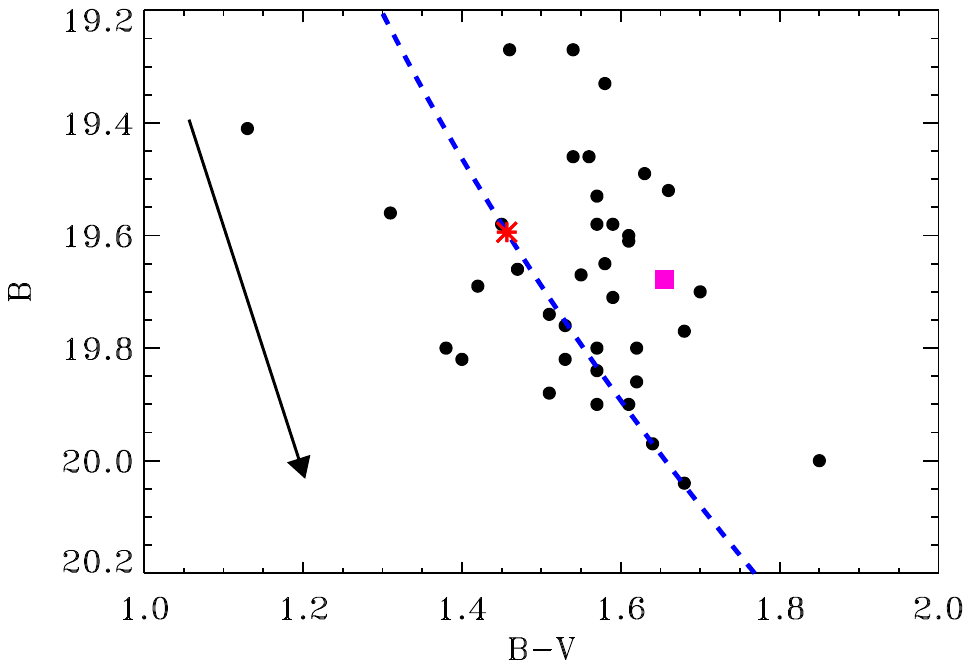}
\includegraphics[trim = 20mm 133mm 30mm 30mm, width=0.32\textwidth]{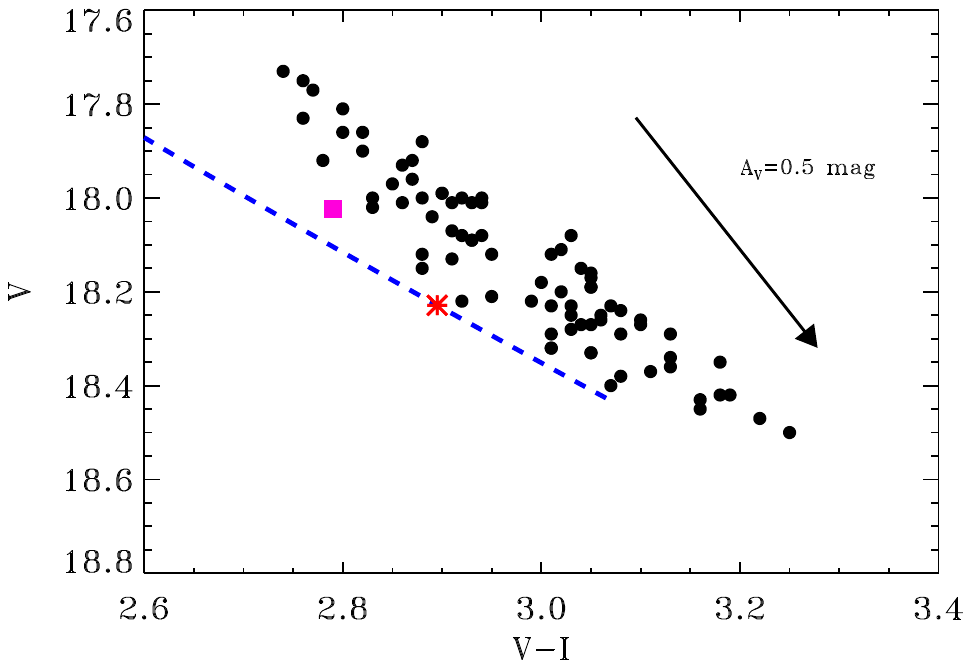}
\includegraphics[trim = 20mm 133mm 30mm 30mm, width=0.32\textwidth]{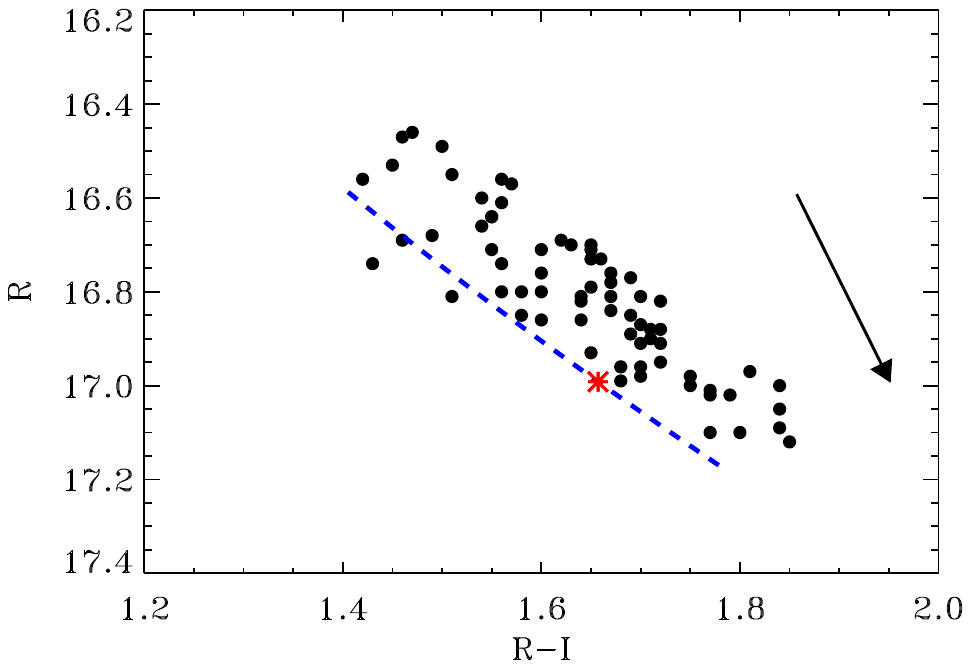}
\caption{BVRI color-magnitude diagrams from the Semkov photometry, compared with synthetic photometry from our fit to SDSS spectra, varying only the accretion continuum (dashed blue line).  The red stars are show the photometry from our best-fit SDSS model.  Each panel also includes an extinction vector for $A_V=0.5$ mag.  The purple square is single-epoch BVI photometry from \citet{guieu09}.}
\label{fig:cmds}
\end{figure*}

The Semkov photometry shows significant variability, with standard deviations of 0.19 in B (ignoring the one extreme outlier) to 0.08 in I (Figure~\ref{fig:lightcurve}).  The strong correlation between source brightness and color, as well as the variability patterns, are more consistent with accretion variability than extinction variability.  Figure~\ref{fig:cmds} shows BVRI photometry from our best-fit model of SDSS photometry, compared with the observed color-magnitude diagram from Semkov photometry.  The model photometry is calculated by scaling only the accretion continuum while keeping the photospheric emission, disk emission, and extinction fixed.

In each case, the accretion vector accurately reproduces the slope of the color-magnitude relationship, while the extinction vector points in a different direction, especially in non-adjacent bands (e.g., V-I).  We conclude that the photometric variability is dominated by accretion and that extinction variability is negligible.
The synthetic photometry is offset from the measured photometry, a consequence of either a different zero point or filter curves.

Because of the offset between the observed and measured photometry, SED fits to the Semkov photometry are considered unreliable.  However, SED fits to the averaged BVRI photometry from the faintest five (V+R) epochs would lead to similar acceptable parameter space for veiling and extinction (as seen in Figure~\ref{fig:contour}), with a long tail to high veiling that could be discarded by relying on the lack of strong veiling in the red spectrum.  For stars without large extinction changes, use of the faintest epochs is important to minimize contamination from accretion.

The full range of B band photometry covers accretion luminosities
rates that are variable by a factor of 5 (maximum/minimum), from 0.012--0.057 $L_\odot$ (see \S 4 for accretion values from SDSS).  The average accretion luminosity inferred from Semkov photometry is consistent with that from SDSS photometry.
The low end of the range is sensitive to the contribution by the photosphere.   If the best-fit to the SDSS spectrum had a lower veiling, the inferred accretion variability would increase because the photosphere would be a higher fraction of the $B$-band magnitude.

\subsection{An updated distance for HBC 722}
We adopt a distance to HBC 722 of $782\pm5$ pc, by measuring the weighted mean Gaia DR3 parallax of 15 Young Stellar Objects from \citet{armond11}, excluding HBC 722, and with parallax uncertainties of less than 1\% \citep{gaiadr3}.  This distance is consistent with but slightly larger than previous Gaia-based measurements to the Pelican Nebula and the broader North America Nebula \citep{bhardwaj19,fang20,kuhn20,kerr23}.

\section{Stellar and Accretion Properties} \label{sec:properties}

Our spectral type of M$3.3\pm0.4$ and temperature $3350\pm75$ K refines previous estimates.   \citet{fang20} used a low S/N spectrum from 5100--7700 \AA\ to estimate a spectral type of M4.6, after accounting for veiling.  Before the veiling correction, the spectrum was assessed as K6.6, similar to the K7/M0 found in the 4300--6700 \AA\ spectrum of \citet{ck79}.  Our high S/N red spectrum includes the prominent TiO bands, allowing for a more accurate spectral type.

The luminosity of HBC 722 is $0.53$ L$_\odot$, calculated from the absolute $J$-band magnitude, after correcting for the distance of 782 pc, extinction $A_V=1.45$ mag, bolometric correction of 1.86 mag from \citet{pecaut13}, and minor contributions from dust emission and accretion.
This luminosity and temperature lead to a radius estimate of 2.15 R$_\odot$.  The estimated mass is 0.39 M$_\odot$ and age 0.51 Myr for the 51\% spot models of \citet{somers20} and 0.24 M$_\odot$ and age  0.15 Myr for the 0\% spot models. The 51\% spot models are adopted here based on the importance of spots at young ages \citep{flores22,cao22,perez25} and consistency with a few accurate dynamical masses in this SpT range \citep{pegues21}.

We estimate an accretion luminosity of $0.034$ L$_\odot$, corresponding to $7\times10^{-9}$ M$_\odot$ yr$^{-1}$, from the SDSS photometry, for the mass of 0.39 M$_\odot$ and assuming a magnetospheric truncation radius of 5~$R_\star$, following the convention established by \citet{gullbring98}. 
 This accretion rate is  about one order of magnitude higher than the average accretion rate for the spectral type, though consistent with the strongest accretors in the distribution \citep{manara23}.
The range of $B$-band variability of 19.2-20.0 from \citet{semkov12} is well reproduced with accretion variability of a factor of $\sim 5$ (maximum/minimum) and a standard deviation of 0.16 dex.  The accretion variability is consistent with measurements of variability for other classical T Tauri stars \citep[see, e.g.][]{venuti21,herczeg23}.

The accretion luminosities inferred from photometry are consistent with the accretion luminosity of $\sim 0.030$ L$_\odot$ from the H$\alpha$ line luminosity and an upper limit of $0.08$ L$_\odot$ for the \ion{He}{1} $\lambda6678$ nondetection, as estimated from correlations provided by \citet{alcala17}.  The \ion{Ca}{2} infrared triplet lines lead to an accretion luminosity of 0.11 L$_\odot$, three times brighter than expected.  The brightness of the infrared triplet lines may be caused by an overabundance of Ca when compared to other accretion flows, a lower temperature in the accretion flow due to the large stellar radius, or perhaps some unusual inner disk structure that preceded the large outburst.  An overabundance of Ca may be expected at early ages due to rapid radial drift \citep[see simulations of Fe by][]{huhn23}.  However, a dependence on abundance has only been detected for accretion flows that are underabundant in Ca, in a sample that lacked such a high ratio of Ca to H$\alpha$ flux \citep{micolta24}.  The blue pre-outburst spectrum of V1057 Cyg had strong \ion{Fe}{1} lines, perhaps for the same reason \citet{herbig58}.

\section{Discussion and Conclusions} \label{sec:conclusions}

Despite the recent discovery of many new FUors, HBC 722 is still unique in having a pre-outburst spectrum with measurable stellar properties.  PR Ori B, reported to be a recent FU Ori outburst \citep{contreras24}, has a pre-outburst 
 infrared spectral type but complicated optical emission \citep{reipurth18}.  V1057 Cyg has a pre-outburst blue spectrum that shows strong lines, including in some H Balmer and \ion{Fe}{2} lines, but no spectral type could be measured \citep{herbig58,herbig77}.  The pre-outburst spectrum of HBC 722 allows us to  measure the pre-outburst properties.  We first measure a spectral type of M3.3$\pm$0.4 from our previously unpublished spectrum, obtained $\sim 7.5$ years prior to the eruption. We then apply that constraint to pre-burst SDSS photometry to measure the stellar luminosity, extinction, and accretion rate and separately to the pre-burst Semkov photometry to assess accretion variability.

The spectral type indicates that HBC 722 has a mass of 0.24-0.39\,M$_\odot$ (depending on the pre-main sequence model), consistent with expectations from line broadening in a viscous disk specifically for HBC 722 (\citealt{carvalho24}, who fixed the mass to 0.2\,M$_\odot$), and more generally for other FU Ori objects \citep[e.g.][]{zhu09,rodriguez22}.  Disk rotation from mm lines should provide tight constraints on the stellar mass but has been difficult to measure for FUors due to the small disk sizes, complicated environments, and distance \citep[e.g.][]{kospal21,weber25}.

The accretion rate during the outburst is $10^{-4}$\,M$_\odot$~yr$^{-1}$ \citep{carvalho24}, or a factor of $15,000$ times higher than the accretion rate of $\sim 7\times10^{-9}$ M$_\odot$ yr$^{-1}$ before the outburst.  This increase is consistent with expectations from population studies but is the first time it has been possible to calculate for an individual object.  \citet{carvalho24} found a higher progenitor accretion rate, adjusted here\footnote{The accretion rate in \citet{carvalho24} is adjusted from $7.8\times10^{-8}$ M$_\odot$ yr$^{-1}$ to account for the mass, radius, and distance adopted here.} to $1.3\times10^{-7}$\,M$_\odot$~yr$^{-1}$,
from the H$\alpha$ emission in the \citet{fang20} spectrum.  This discrepancy is dominated by 
the lower extinction measured in this paper.  The high accretion rate rules out the possibility that, prior to the outburst, gas was accumulating at the inner disk but prevented from accreting onto the star.

Our analysis reveals two somewhat peculiar traits of the progenitor of the HBC 722 outburst.  First, the radius is about 2 times larger than the radii of stars with similar spectral types in Lupus \citep[see the table with distance updates from Gaia DR1 in][]{alcala19}, with only one exception (Sz 74).  The large radius is consistent with youth (the HR diagram based age estimates are 0.15-0.51 Myr) or perhaps with radius inflation from periods of past accretion \citep{stahler83,hartmann11}.  Second, the \ion{Ca}{2} IR triplet is three times stronger than expected for the accretion rate, which may be caused by emission from outflows, an inflated inner disk, or a cooler accretion flow.

Our fits to photometry lead to extinction measurements of $A_V=1.45\pm0.3$ and are robust to reasonable changes in assumptions.
In contrast, the current outburst has a much higher extinction.  Full spectral fits led to an extinction of $A_V=2.5$ mag (\citealt{carvalho24}, see also $A_V=3.1$ from \citealt{rodriguez22}), while near-IR colors lead to $A_V=3.7$ \citep{connelley18}.

The pre-outburst extinction cannot be reconciled with the outburst extinction measurement. 
Our intuition was that any change should be in the opposite direction, that the extinction could have decreased during the outburst due to strong winds  \citep[e.g.][]{croswell87}.  The increase in extinction during the outburst may be caused by dusty winds \citep[see for instance speculation about the fade of the non-outburst of RW Aur A by][]{petrov15}.  Simulations by \citet{kadam25} indicate that MHD winds from outbursting stars may be more dusty than the winds emitted during quiescence, though they are still dust-depleted compared to the ISM.
\citet{carvalho24} confirm that HBC 722 has a strong, cool wind and an evolving disk. Alternatively, the outburst may increase the geometrical thickness of the disk, so that our line of sight to the emission region (the innermost disk) passes through the disk at larger radii.  Large decreases in extinction has been measured previously during some large outbursts, including that of V1647 Ori \citep{aspin11}, V346 Nor \citep{kospal17}, Gaia 19ajj \citep{hillenbrand19}, and the bonafide FU Ori outburst PTF10nvg \citep{hillenbrand13,kospal13}.  However, increased extinction from quiescence to an FU Ori outburst has not been previously measured, and extinction for HBC 722 has been assumed to be constant \citep[e.g.][]{miller11,carvalho24}.  The Semkov photometry demonstrates that extinction variability was also negligible prior to the outburst.   We conclude that analysis of pre-burst photometry of FU Ori objects cannot assume that the extinction will remain unchanged after an outburst, though the change in extinction may also provide us with an opportunity to assess changes in disk or wind structure.

The fitting approach developed here has implications for future analyses of YSOs, including estimating stellar parameters of outbursts before the burst.  Without the u-band, the acceptable parameter space includes a long tail to higher accretion, higher extinction, and weaker photometric emission.  Moreover, fits to photometry should include accretion, otherwise the stellar parameters are poorly constrained \citep[see also][]{piscarreta25}.
 For accreting young stars with long-term monitoring, such as from LSST \citep{bonito23} or WFST, the stellar parameters are likely best established from faint epochs.

\begin{acknowledgements}

We thank the anonymous referee for a careful read and identifying concerns that helped to improve the paper.
We thank Tina Armond for participating in the observations and Colin Aspin for reducing the spectra. GJH thanks Min Fang and Adolfo Carvalho for valuable discussions.  
GJH is supported by National Key R\&D
program 2022YFA1603102 from the Ministry of
Science and Technology (MOST) of China and by the general
grant 12173003 from the National Natural Science Foundation
of China.

\end{acknowledgements}

\bibliographystyle{apj}
\bibliography{ref}
\end{CJK*}

\end{document}